\documentclass[epsfig,twocolumn,showpacs,preprintnumbers,amsmath,amssymb]{revtex4}
\usepackage{graphicx}% Include figure files
\usepackage{dcolumn}% Align table columns on decimal point
\usepackage{bm}% bold math

\begin{document}

\preprint{NSF-KITP-05-15}

\title{Theory of the critical current in two-band superconductors
with application to MgB$_2$}
\author{E.J. Nicol}
\email{nicol@physics.uoguelph.ca}
\affiliation{Department of Physics, University of Guelph,
Guelph, Ontario, N1G 2W1, Canada}
\author{J.P. Carbotte}
\email{carbotte@mcmaster.ca}
\affiliation{Department of Physics and Astronomy, McMaster University,
Hamilton, Ontario, L8S 4M1, Canada}
\date{\today}

\begin{abstract}
Using a Green's function formulation of the superfluid current $j_s$,
where a momentum $q_s$ is applied to the Cooper pair, we have
calculated $j_s$ as a function of $q_s$, temperature, and
impurity scattering for a two-band superconductor. We consider
both renormalized BCS 
and full strong-coupling Eliashberg theory. There are
two peaks in the current as a function of $q_s$
due to the two energy scales for the gaps 
and this can give rise to non-standard behavior for
the
critical current. The critical current $j_c$, which is given as the
maximum in $j_s$, can exhibit a kink as a function of temperature
as the maximum is transferred from one peak to other. Other
temperature variations are also possible and the universal BCS
behavior is violated. The details
depend on the material parameters of the system, such as the amount
of coupling between the bands, the gap anisotropy, the Fermi
velocities,
and the density of states of each band. The Ginzburg-Landau relation
between $j_c$, the penetration depth $\lambda_L$ and thermodynamic 
critical field $H_c$, is modified. Using Eliashberg theory with the 
electron-phonon spectral
densities given from bandstructure calculations,
we have applied our calculations
for $j_s$ and $j_c$  to the case
of MgB$_2$ 
and find agreement with experiment.

\end{abstract}

\maketitle
% body of paper here

\section{Introduction}

Two-band superconductivity was first proposed in the late 1950's
and
further studied in the 1960's and 1970's as a possible explanation for
understanding superconductivity in s- and d- band metals. However, it
was not until the discovery of superconductivity at relatively high 
temperature in MgB$_2$ in 2001, that the most promising
example of two-band superconductivity was found. After nearly four years of
effort and the development of an extensive literature based on this
material, it is now evident that this compound is an electron-phonon
mediated superconductor, but with two distinct bands and hence two
energy gaps. Indeed, this is firmly established from detailed
bandstructure calculations that provide the electron-phonon spectral
densities
which, when in turn are used in a two-band Eliashberg formalism, give
rise  to predictions which are in accord with experiment. In fact,
the picture that emerges is one of excellent agreement with the data,
with the match being as good as is found between theory and
experiment
in conventional one-band superconductors\cite{carbotte}. A detailed review and
summary of the level of agreement found for the thermodynamic
properties,
BCS ratios, penetration depth and temperature-dependent energy gaps
is found in Ref.~\cite{nicolmg}. With such an extensive comparison
between rigorous Eliashberg theory and experiment completed,
it is now appropriate to move on to the transport properties where less
theoretical
work has been done.

In this paper, we consider the critical current which is of
fundamental
interest  in the discussion of transport. While most technological
applications
 focus on the bulk critical current density, which requires
understanding of flux lattice pinning, etc., here we are
presenting the critical current in thin films. 
This geometry  is more relevant to understanding issues
associated with the fundamental superconducting state, rather than
focussing on
material
details that lead to vortex pinning, etc., and hence can be expected
to provide further insight  into the nature of multiband
superconductivity in these materials. We note in passing,
at this point, that
some authors 
use the term depairing current for the
critical current although, technically, the depairing
current is the current at which the first Cooper pair is
broken and the gap decreases, 
and the critical current is the maximum current possible,
which is greater than the depairing current. In simple one-band
s-wave BCS theory the values are almost the same and hence there is
the tendency to use the terms interchangeably. Here, these two
quantities can be quite different and so we will use the term
critical current
to refer to the global maximum obtainable in the current 
density.

The full temperature-dependence of the 
critical current $j_c$  was first studied in conventional one-band
s-wave BCS superconductors by Rogers\cite{rogers} and
Bardeen\cite{bardeen}. 
Further work following on this
was done by
Parameter\cite{parameter} for the clean limit,
and by Maki\cite{makiold,maki} for the dirty limit, 
where the latter author introduced a
description of the superfluid current density using the Green's
function
formalism. A study of the effect of of a current on the
quasiparticle states density of states of a superconductor
was done by 
Fulde\cite{fulde}. The most general work 
was that of Kupriyanov and Lukichev\cite{lukichev},
who calculated the full temperature dependence
of the critical current in BCS theory for arbitrary impurity scattering, 
which returned the usual
Ginzburg-Landau results for the critical current near
$T_c$\cite{tinkham}.
Further considerations led to an initial attempt to include strong
inelastic
scattering\cite{lemberger} followed by a full Eliashberg calculation
and study   of strong coupling effects for conventional
electron-phonon
superconductors done by ourselves\cite{nicoljc}. Thus, the formalism
and basic one-band results have been well-established in  both the
weak-coupling
BCS limit and strong-coupling Eliashberg regime, at all temperatures
and
for arbitrary impurity scattering.

More recently there has been a renewed theoretical effort on the critical
current as applied to exotic gaps, which are anisotropic, such as
d-wave\cite{ting,hykeenodal} and f-wave\cite{hykeenodal,hykee}. The
two-band case  is, in some sense, a highly anisotropic system and
hence the study of the critical current in this instance  also
complements
these recent  works.

With regard to MgB$_2$, specifically, there has been some recent
experimental
work by Kunchur\cite{kunchur}, where the critical current has
been measured in these systems. On the
theoretical
side, Koshelev and Golubov\cite{koshelevjc} have gone beyond
Ginzburg-Landau to use the Usadel equations to study this system and
find interesting effects such as a kink in the temperature-dependence
of
the c-axis critical current that reflects the underlying two-band nature.
This type of feature is observed in other properties such as the
penetration
depth which also shows an inflection point in the temperature
dependence, however, these effects may disappear with increased
interband coupling. Finally, a more recent work has been done on
the topic of the non-linear term  in the superfluid current 
in MgB$_2$ and its effect on the nonlinear microwave response,
which is relevant for device applications\cite{dahm}.

In our work here, we use the general Green's function formulation
presented
by Maki\cite{maki} where the Cooper pairs are given a finite center
of mass momentum $q_s$. This $q_s$ leads to a boost
in the quasiparticle energy (sometimes
called a Doppler shift). The superfluid velocity $v_s=q_s/m$ is
taken as uniform in the case of discussing currents in thin films
or wires,
unlike the Doppler shift effect in the context of
superfluid currents in the vortex state, where $v_s$ varies with
spatial position. This latter topic
has been the subject of a large number of papers in the recent
literature relating to
d-wave superconductors\cite{vekhter}.

A very fine physical description of the effect of this Doppler shift
on the quasiparticle density of states and the resulting effect on
the
superfluid density is presented in a paper by Xu et al.\cite{sauls},
where they consider the non-linear Meissner effect in d-wave
superconductors
from a starting point that discusses the basics of the superfluid
current
density.

Thus, with a body of literature developing on the topic of critical
currents
and superfluid currents, it is of interest to elucidate the features
in the 
critical current due to two-band superconductivity, where unusual
effects can occur due to a transfer from one band to the other
of the dominant contribution to superfluid current and hence
the critical current.

Our paper is structured in the following manner. In Section~II, we
introduce the basic theoretical equations which we have evaluated.
We then examine, in Section~III, results for renormalized BCS theory
near $T_c$ and at $T=0$ and apply these results to MgB$_2$.
In Section~IV, we present our Eliashberg results for
finite temperature and with impurities, evaluating the critical
current for MgB$_2$ and other model parameters. We also compare
our results to the data for MgB$_2$. Finally, we form
our conclusions in Section~V. 

\section{Theory}

While various approaches have been used in the past for
calculating the superfluid current, we will use the Green's function
method as presented by Maki\cite{maki}. Thus the superfluid current
density $\vec j_s$  is given in terms of finite temperature
Green's functions as:
\begin{equation}
\vec{j}_s=\frac{eT}{m}\sum _{n=-\infty}^\infty\int \frac{d^3p}{(2\pi)^3}{\rm Tr}[
\vec{p}\,G_{\omega_n,\vec{q}_s}(\vec{p})]\quad ,
\label{eq:js1}
\end{equation}
where $e$ is the electron charge,  $m$, the electron mass, $\vec p
= m\vec v$
is the electron momentum, and $T$ is
the temperature. The Green's function is given in terms of
the Matsubara representation by
\begin{equation}
G_{\omega _n,\vec q_s}(\vec p)=[i\tilde\omega(n)-\vec v\cdot\vec q_s+
\xi\rho _3+\tilde\Delta(n)\rho _{_1}\sigma _{_2}]^{-1},
\label{eq:greensfn}
\end{equation}
where $\xi=p^2/2m-\mu$, $\mu$ is the chemical potential, and the $\rho_i$
and
$\sigma_i$ are the Pauli matrices for the particle-hole and spin
spaces,
respectively. The  $\vec q_s$ is the applied superfluid momentum and
it results in a shift of the quasiparticle energies, given
to first order in $q_s$ by $\vec v\cdot\vec q_s$. 
Here,  $\tilde\Delta(n)$  and $\tilde\omega(n)$  are the imaginary 
axis renormalized 
superconducting order parameter and renormalized 
Matsubara frequencies, respectively, and are defined below.

The integral over energy $\xi$ for the combined
Eqns.~(\ref{eq:js1}) and (\ref{eq:greensfn}) can be done and the 
result, generalized to two bands, is given in terms of a sum
of two partial currents $j_{s1}$ and $j_{s2}$,
for the first and second band, respectively:
\begin{equation}
j_s= j_{s1}+j_{s2}
\end{equation}
with
\begin{equation}
{j}_{si}=\frac{3en_i}{mv_{Fi}}\pi T\sum_{n=-\infty}^{+\infty}\int_{-1}^1dz\,
\frac{i(\tilde\omega _i(n)-is_iz)z}{\sqrt{(\tilde\omega _i(n)-is_iz)^2+
\tilde\Delta^2 _i(n)}},
\label{eq:js2}
\end{equation}
where $s_i=v_{Fi}q_s$,
$n_i$ is the electron density and
$v_{Fi}$ is the Fermi velocity of the $i$'th band ($i=1$,2).
The nonlinear
Eliashberg equations for 
$\tilde\Delta _i(n)=Z_i(n)
\Delta_i(n)$ and $\tilde\omega _i(n)=Z_i(n)\omega _n$
have been generalized to two bands and with the
inclusion of the effect of the $q_s$, they are given as\cite{nicolmg,nicoljc}:
\begin{eqnarray}
\tilde\Delta_i(n) &=& \pi T\sum_m\sum_j[\lambda_{ij}(m-n)
-\mu^*_{ij}(\omega_c)\theta(\omega_c-|\omega_m|)]\nonumber\\
&\times&\int_{-1}^{1}\frac{dz}{2}
\frac{\tilde\Delta_j(m)}{\sqrt{(\tilde\omega_j(m)-is_jz)^2+\tilde\Delta_j^2(m)}}\nonumber\\
&+& \pi\sum_j(t^+_{ij}-t^-_{ij})\int_{-1}^{1}\frac{dz}{2}
\frac{\tilde\Delta_j(n)}
{\sqrt{(\tilde\omega_j(n)-is_jz)^2+\tilde\Delta_j^2(n)}}\nonumber\\
\label{eq:Del}
\end{eqnarray}
and
\begin{eqnarray}
\tilde\omega_i(n) &=& \omega_n+\pi T\sum_m\sum_j
\lambda_{ij}(m-n)\nonumber\\
&\times&\int_{-1}^{1}\frac{dz}{2}
\frac{\tilde\omega_j(m)-is_jz}{\sqrt{(\tilde\omega_j(m)-is_jz)^2+\tilde\Delta_j^2(m)}}\nonumber\\
&+& \pi\sum_j(t^+_{ij}+t^-_{ij})\int_{-1}^{1}\frac{dz}{2}
\frac{\tilde\omega_j(n)-is_jz}
{\sqrt{(\tilde\omega_j(n)-is_jz)^2+\tilde\Delta_j^2(n)}},\nonumber\\
\label{eq:Z}
\end{eqnarray}
where $t^+_{ij}=1/(2\pi\tau^+_{ij})$ 
and  $t^-_{ij}=1/(2\pi\tau ^-_{ij})$ are the 
ordinary and paramagnetic impurity scattering rates, respectively,
and the electron-phonon kernels $\alpha^2_{ij}F(\Omega)$ with 
phonon energy $\Omega$  enter through
\begin{equation}
\lambda_{ij}(m-n)\equiv 2\int^\infty_0\frac{\Omega\alpha^2F_{ij}(\Omega)}{
\Omega^2+(\omega_n-\omega_m)^2}d\Omega.
\end{equation}
Here, the $n$  indexes the $n$'th  Matsubara frequency $\omega_n$,
with $\omega_n=(2n-1)\pi T$, where $n=0,\pm 1,
\pm 2,\cdots$. The $\mu^*_{ij}$ are Coulomb repulsions,
with a high energy cutoff $\omega_c$ 
usually taken to be about six to ten 
times the maximum phonon frequency. As written,
Eqns.~(\ref{eq:js2})-(\ref{eq:Z})
are for three dimensions with isotropic Fermi surface. For
general anisotropic Fermi surface, a Fermi surface integral would
remain, which would need to be done numerically\cite{dahm}.
This goes beyond the scope of this
paper, however, such details are not expected to change the qualitative
features of our results but rather lead to quantitative changes only.

In the following sections, we present both Eliashberg results and renormalized
BCS (RBCS) results, the latter of which can be quite successful at capturing
the
essential features of the Eliashberg calculations without the full
numerical complications of the more sophisticated
calculations
even for two-band models\cite{nicolmg}. 
To develop the RBCS results we use
the two-square-well approximation or $\lambda^{\theta\theta}$
model written for two-bands (for details see Ref.~\cite{nicolmg}), 
where now
\begin{equation}
\Delta_i(T)=\frac{\pi T}{Z_i}\sum_{m,|\omega_m|<\omega_D}\sum_j
\int^1_{-1}\frac{dz}{2}
\frac{[\lambda_{ij}-\mu^*_{ij}]\Delta_j(T)}{\sqrt{(\omega_m-i\bar s_jz)^2+\Delta_j^2}}.
\label{eq:rbcsdel}
\end{equation}
In this expression,  $\bar s_j=s_j/Z_j$,
\begin{equation}
\lambda_{ij}=2\int^\infty_0
\frac{\alpha^2F_{ij}(\Omega)}{\Omega}d\Omega
\end{equation}
and
\begin{equation}
Z_i=1+\sum_j\lambda_{ij},
\label{eq:rbcsZ}
\end{equation}
with $\omega_D$ taken to
represent either the Debye frequency or some other
characteristic energy scale representing the phonons in the system. 
Likewise, the current  from Eq.~(\ref{eq:js2}) can be reduced to
\begin{equation}
{j}_{si}=\frac{3en_i}{mv_{Fi}}\pi T\sum_{n=-\infty}^{+\infty}\int_{-1}^1dz\,
\frac{i(\omega_n-i\bar s_iz)z}{\sqrt{(\omega _n-i\bar s_iz)^2+
\Delta^2 _i(T)}}.
\label{eq:rbcsjs}
\end{equation}
Equations~(\ref{eq:rbcsdel}) and (\ref{eq:rbcsjs}) will be used in the
next section to develop analytic results for comparison with full
numerical calculation.

\section{Renormalized BCS Results}

\subsection{RBCS near $T_c$}

In order to evaluate the critical current near $T_c$, we need to
know the effect of the current on the gap near $T_c$. In this
limit, both $\Delta$ and $q_s$ will be small.
After long but straightforward algebra, to first order in $(1-t)$
where $t=T/T_c$ is the reduced temperature, the
gap equation of Eq.~(\ref{eq:rbcsdel}) reduces to the form
\begin{equation}
1=\bar\lambda_{11}F_1+\frac{\bar\lambda_{12}\bar\lambda_{21}}{1-\bar\lambda_{22}F_2}F_1F_2+\Delta_1^2G(T_c)H_1,
\label{eq:bcs1}
\end{equation}
with a second equation obtained from Eq.~(\ref{eq:bcs1}) by switching
indices $1\leftrightarrow 2$. Here
\begin{equation}
\bar\lambda_{ij}=\frac{\lambda_{ij}-\mu^*_{ij}}{Z_i}
\end{equation}
and
\begin{equation}
F_i\equiv F_i(T)={\rm ln} \biggr(\frac{1.13\omega_D}{T}\biggr)+\frac{2}{3}G(T)v_{Fi}^{*2}q_s^2,
\label{eq:bcs2}
\end{equation}
where $v_{Fi}^*\equiv v_{Fi}/(1+\lambda_{ii}+\lambda_{ij})$
and  $G(T)=-7\zeta(3)/8(\pi T)^2$. For 
$\Delta_1(T)\to 0$ as $T\to T_c$ and with no superfluid momentum $(q_s=0)$,
we recover the equation for the critical temperature (symmetric in 
$1\leftrightarrow 2$)\cite{nicolmg}:
\begin{equation}
1=(\bar\lambda_{11}+\bar\lambda_{22})g+(\bar\lambda_{12}\bar\lambda_{21}-
\bar\lambda_{11}\bar\lambda_{22})g^2,
\label{eq:bcs3}
\end{equation}
with $g=\ln(1.13\omega_D/T_c)$. Throughout this paper, we will find it
instructive to consider several simplifying limiting cases. The completely
decoupled case corresponds to $\lambda_{12}=\lambda_{21}=0$ and the
assumption that $\bar\lambda_{11}>\bar\lambda_{22}$. In this case, the first
channel determines the $T_c$, which is given by
$1.13\omega_D\exp(-1/\bar\lambda_{11})$. As an aside, we note that
the small separable $a^2$ anisotropy model studied extensively in the
older literature\cite{aniso}
 maps on to this model if $\bar\lambda_{11}=\bar\lambda (1+a)^2/2$,
$\bar\lambda_{22}=\bar\lambda(1-a)^2/2$ and 
$\bar\lambda_{12}\bar\lambda_{21}=
\bar\lambda(1-a^2)/2$, with the anisotropy $a^2$ assumed small. In this case,
Eq.~(\ref{eq:bcs3}) simplifies greatly since  
$\bar\lambda_{12}\bar\lambda_{21}
=\bar\lambda_{11}\bar\lambda_{22}$ so that 
$g=1/(\bar\lambda_{11}+\bar\lambda_{22})=1/[\bar\lambda(1+a^2)]
=(1+\lambda)/[\lambda(1+a^2)]$ which gives 
$T_c=1.13\omega_D \exp(-(1+\lambda)/[\lambda(1+a^2)])$.
This is the well-known result that
anisotropy increases the value of the critical temperature
over its isotropic value ($a^2=0$).

Returning to Eq.~(\ref{eq:bcs1}) which gives the gap just below $T_c$,
we need to specify $H_i$. It is given (after more
lengthy algebra) as
\begin{equation}
H_1=\bar\lambda_{11}+
\frac{g\bar\lambda_{12}\bar\lambda_{21}}{1-g\bar\lambda_{22}}\biggl\{1+\frac{g^2\bar\lambda_{21}^2}
{(1-g\bar\lambda_{22})^3}\biggr\}
\label{eq:bcs4}
\end{equation}
and an equivalent expression for $H_2$ is obtained
from Eq.(\ref{eq:bcs4}) by switching indices $1\leftrightarrow 2$.
Solving for $\Delta_1^2(t)$, we find
\begin{equation}
\Delta_1^2(t)=
-\frac{(1-t)}{G(T_c)}\frac{1}{\chi_1}-\frac{2}{3}q_s^2\frac{1}{\chi_1^\prime},
\label{eq:bcs5}
\end{equation}
with
\begin{equation}
\frac{1}{\chi_1}=\frac{1}{H_1}\biggl[\bar\lambda_{11}+
\frac{\bar\lambda_{12}\bar\lambda_{21}}{1-g\bar\lambda_{22}}
\biggl(2g+\frac{g^2\bar\lambda_{22}}
{1-g\bar\lambda_{22}}\biggr)\biggr]
\label{eq:bcs6}
\end{equation}
and
\begin{equation}
\frac{1}{\chi_1^\prime}=\frac{1}{H_1}\biggl[\bar\lambda_{11}v_{F1}^{*2}+
\frac{g\bar\lambda_{12}\bar\lambda_{21}}{1-g\bar\lambda_{22}}\biggl(v_{F1}^{*2}
+v_{F2}^{*2}+\frac{g\bar\lambda_{22}v_{F2}^{*2}}
{1-g\bar\lambda_{22}}\biggr)\biggr].
\label{eq:bcs7}
\end{equation}
Likewise, the gap
$\Delta_2(t)$, is formed from the above three equations
by exchanging $1\leftrightarrow 2$. When $q_s=0$, these equations
properly reduce to those obtained previously by Nicol and 
Carbotte\cite{nicolmg}. In the limit of decoupled bands, $1/\chi_1 =1$
and $1/\chi^\prime_1=v_{F1}^{*2}$. Some care is required when treating
the equivalent equations for $1/\chi_2$ and $1/\chi^\prime_2$ because
the combination $(1-g\bar\lambda_{22})$ is replaced by $(1-g\bar\lambda_{11})$,
which is zero. Nevertheless, we can show $1/\chi_2=
1/\chi^\prime_2=0$, which is expected on physical grounds as the second
band does not contribute near $T_c$, where $\Delta_2^2(t)=0$
for $t\to 1$, and  $\Delta_1^2(t)=(1-t)8(\pi T_c)^2/[7\zeta(3)]$, a
well-known result. We can also recover other well-known results for
the small separable anisotropy model defined previously. Note first that
when the Fermi velocities are equal, $1/\chi_1^\prime$ in Eq.~(\ref{eq:bcs7})
reduces to $v_F^{*2}/\chi_1$, with $1/\chi_1$ given by Eq.~(\ref{eq:bcs6}).
Further, $1/\chi_1=(1+a)^2(1-5a^2)$ and $1/\chi_2=(1-a)^2(1-5a^2)$ which
leads to (Eq.~(\ref{eq:bcs5}))
\begin{equation}
\Delta_1^2(t)=(1+a)^2(1-5a^2)\biggl[(1-t)\frac{8(\pi T_c)^2}{7\zeta(3)}-
\frac{2}{3}q_s^2v_{F}^{*2}\biggr]
\label{eq:bcs8}
\end{equation}
and an identical equation for the second gap with the first factor of
$(1+a)^2$ replaced by $(1-a)^2$. For $q_s=0$, Eq.~(\ref{eq:bcs8}) reduces
to the known result that the average gap 
\begin{equation}
\Delta_0^2(t)=(1-5a^2)(1-t)\frac{8(\pi
T_c)^2}{7\zeta(3)},
\label{eq:bcs9}
\end{equation}
with $\Delta_1(t)=\Delta_0(t)(1+a)$ and $\Delta_2(t)=\Delta_0(t)(1-a)$.
For the isotropic case (equivalent to one band) $a^2=0$ and
Eq.~(\ref{eq:bcs8}) reduces to 
\begin{equation}
\Delta^2(t)=(1-t)\frac{8(\pi T_c)^2}{7\zeta(3)}-\frac{2}{3}q_s^2v_F^{*2},
\label{eq:bcs10}
\end{equation}
as is well known.

The expression for the contribution to the current from the $i$'th
band is given in renormalized BCS by Eq.~(\ref{eq:rbcsjs}).
Expanding near $T=T_c$ for small $q_s$ and $\Delta$ gives
\begin{equation}
j_{si}(t)=\frac{n_ie}{m_iv_{Fi}}
\frac{7\zeta(3)}{4(\pi T_c)^2}\Delta_i^2(t)v_{Fi}^*
q_s.
\label{eq:bcs12}
\end{equation}
The total current is the sum over both bands and can be written as
\begin{equation}
j_s(t)=\biggl[\frac{n_1}{m_1^*}\Delta_1^2(t)+\frac{n_2}{m_2^*}\Delta_2^2(t)
\biggr]
\frac{7\zeta(3)}{4(\pi T_c)^2}
eq_s,
\label{eq:bcs13}
\end{equation}
where $m_i^*=m_i(1+\lambda_{ii}+\lambda_{ij})$. Substituting into
Eq.~(\ref{eq:bcs13}) the new expression from Eq.~(\ref{eq:bcs5})
(and equivalent for $1\leftrightarrow 2$) for $\Delta_1^2(t)$, one
obtains two terms. The first proportional to $q_s$ and the second to
$q_s^3$. This gives the total current $j_s$ as a function of $q_s$.
The critical current $j_c(t)$ is obtained as the maximum of $j_s(t,q_s)$
as a function of $q_s$. After finding the extremum value of $q_s$,
we obtain for $j_c(t)$
\begin{eqnarray}
j_c(t)&=&\frac{8e\pi T_c}{3\sqrt{7\zeta(3)}}(1-t)^{3/2}\nonumber\\
&\times &\biggl[
\frac{n_1}{m_1^*}\frac{1}{\chi_1}+\frac{n_2}{m_2^*}\frac{1}{\chi_2}\biggr]^{3/2}
\biggl[
\frac{n_1}{m_1^*}\frac{1}{\chi_1^\prime}+\frac{n_2}{m_2^*}
\frac{1}{\chi_2^\prime}\biggr]^{-1/2}.
\label{eq:bcs14}
\end{eqnarray}
We first note that the standard Ginzburg-Landau 
$(1-t)^{3/2}$ temperature dependence of the
critical current remains unmodified in the two-band model. As a 
first check on our expression in Eq.~(\ref{eq:bcs14}), we can recover
the known one-band result when both bands are assumed to be identical
by taking $\lambda_{ij}=\lambda/2$, $v_{F1}=v_{F2}$, and the
value of $n_1=n_2=n_T/2$. Here $n_T$ is the total electron density per unit
volume equal to the sum $n_1+n_2$. We obtain the standard
Ginzburg-Landau
result\cite{lukichev}:
\begin{equation}
j_c(t)=\frac{8en_T}{3mv_F}\biggl(\frac{\pi T_c}{\sqrt{7\zeta(3)}}\biggr)
(1-t)^{3/2}.
\label{eq:bcs15}
\end{equation}
Note that all effective mass renormalization $(1+\lambda)$ factors
have cancelled in Eq.~(\ref{eq:bcs15}). Applying the separable 
gap anisotropy
model to Eq.~(\ref{eq:bcs14}) leads to a modification of Eq.~(\ref{eq:bcs15})
by a multiplicative factor of $(1-4a^2)$. This gap 
anisotropy will decrease $j_c(t)$
over its value for $a=0$. Decoupled bands give the usual expression
of Eq.~(\ref{eq:bcs15}) since the second band makes no contribution near
$T=T_c$, but now $n$, $m$, and $v_F$ are those of the first band only.
In the general case, we need to return to Eq.~(\ref{eq:bcs14}) which
tells
us how material parameters affect the slope of $j_c(t)^{2/3}$ which
varies linearly with $(1-t)$ near $T_c$. If we normalize the current,
as we will in the next section, to a slope of negative one near $T_c$, this
will mean that the current at $T=0$ will vary with material
parameters  in contrast to the one-band BCS curve for which it is
a universal number for the clean limit case that we are discussing here.

A well known result of Ginzburg-Landau theory is that near $T_c$,
$H_c(t)/\lambda_L(t)=(3\pi\sqrt{6}/c)j_c(t)$, where $H_c(t)$ is the
thermodynamic critical field and $\lambda_L(t)$ is the London penetration
depth, both for the case of zero current. In a previous paper\cite{nicolmg}, 
we have given expressions for $H_c(t)$ and $1/\lambda_L^2(t)$ as
$t\to 1$. Using these with Eq.~(\ref{eq:bcs14}), we arrive at the
expression
relating the two-band $H_c$, $\lambda_L$ and $j_c$:
\begin{equation}
\frac{H_c(t)}{\lambda_L(t)j_c(t)}=\frac{3\pi\sqrt{6}}{c}J,
\label{eq:bcs17}
\end{equation}
where $c$ is the velocity of light and $J$ is a complicated 
material-dependent parameter. It is given by
\begin{equation}
J = \sqrt{\biggl(\frac{N_1^*v_{F1}^{*2}}{\chi_1^\prime}+
\frac{N_2^*v_{F2}^{*2}}{\chi_2^\prime}\biggr)
\biggl(\frac{N_1^*}{\chi_1^2}+
\frac{N_2^*}{\chi_2^2}\biggr)}
\biggl[\frac{N_1^*v_{F1}^{*2}}{\chi_1}+
\frac{N_2^*v_{F2}^{*2}}{\chi_2}\biggr]^{-1}
\label{eq:bcs18}
\end{equation}
where we have found it convenient to change from electron density $n_i$,
to density of states at the Fermi surface $N_i(0)\equiv N_i$.
This modifies the usual Ginzburg-Landau relation by a constant factor. 
For the decoupled
case where $1/\chi_2=1/\chi_2^\prime=0$,
$1/\chi_1^\prime=v_{F1}^{*2}$, and
$1/\chi_1=1$, $J$ becomes equal to 1. There is no change
in the Ginzburg-Landau relation and this makes
sense since near $t=1$ we are dealing with the single band (the first
one). 
For the
isotropic gap case, where  $\lambda_{ij}=\lambda/2$ for any $(i,j)$,
from Eqs.~(\ref{eq:bcs6}) and (\ref{eq:bcs7}) we obtain
$1/\chi_1=1/\chi_2=1$ and $1/\chi_1^\prime=1/\chi_2^\prime= (v_{F1}^{*2}
+v_{F2}^{*2})/2$, the average of the squares of the Fermi velocities
in the two bands, which gives
\begin{equation}
J = \sqrt{\frac{(N_1+N_2)(v_{F1}^{*2}+v_{F2}^{*2})}
{2(N_1v_{F1}^{*2}+N_2v_{F2}^{*2})}}.
\label{eq:bcs19}
\end{equation}
For identical bands, this expression gives
1, as it must. This also holds if either the two Fermi velocities
are the same or the two density of states match, but not if both
parameters
for one band differ from those of the other band.
We see that, in general, $J$ can be either greater or less than one
and the Ginzburg-Landau relation ceases to hold. 

For identical bands but with gap anisotropy in the separable small
anisotropy $a^2$ model, $1/\chi_1^\prime = v_F^2/\chi_1$ and 
$1/\chi_2^\prime=v_F^2/\chi_2$, with $1/\chi_1=(1+a)^2(1-5a^2)$
and $1/\chi_2=(1-a)^2(1-5a^2)$ which gives $J=1$ to order $a^2$.
This result is expected since, as is well known, the $a^2$ anisotropy
correction to $H_c(t)$ and $1/\lambda_L(t)$ are both of the form
$(1-2a^2)$ which gives a factor of $(1-4a^2)$ in the ratio 
$H_c(t)/\lambda_L(t)$. This agrees perfectly with the anisotropy factor
we derived earlier for $j_c(t)$. Thus, in this case, the
Ginzburg-Landau
relation will hold.

We wish to comment on the use
of the Ginzburg-Landau relation 
$H_c(t)/\lambda_L(t)=(3\pi\sqrt{6}/c)j_c(t)$ for comparison 
with data at all temperatures, which is done 
by substituting two-fluid model forms for $H_c(t)$ and $\lambda_L(t)$.
The Ginzburg-Landau result should only hold near $T_c$ and indeed
we find that when we form the ratio of $H_c(t)/\lambda_L(t)$ using
the $H_c(t)$ and $\lambda_L(t)$ from the full Eliashberg procedure,
the resulting $j_c(t)$ curve rapidly deviates from 
the Eliashberg calculated $j_c(t)$. The results of using such a 
phenomenological  procedure can give curves which are lower or higher
than the true result, and will not even reproduce the qualitative feature
of a kink in the $j_c(t)$ that we have found in the two-band calculation.
We conclude that the Ginzburg-Landau procedure should only be used near
$T_c$ as is appropriate to its derivation and no reliable use can be
made of it phenomenologically for lower temperatures.

\subsection{RBCS for $T=0$}

We now turn to low temperature and examine results of RBCS at
$T=0$.
At zero temperature, the renormalized BCS gap equation (\ref{eq:rbcsdel})
takes on the form
\begin{eqnarray}
\Delta_1&=&\bar\lambda_{11}\Delta_1 {\cal G}(\Delta_1,\bar s_1)
+\bar\lambda_{12} \Delta_2\, {\cal G} (\Delta_2,\bar s_2)\\
\Delta_2&=&\bar\lambda_{21}\Delta_1 {\cal G}(\Delta_1,\bar s_1)
+\bar\lambda_{22} \Delta_2\, {\cal G} (\Delta_2,\bar s_2)
\end{eqnarray}
where the barred quantities are renormalized by the factor of $Z_i$ of
Eq.~(\ref{eq:rbcsZ}) and
\begin{eqnarray}
{\cal G}(\Delta,\bar s) &=& \ln\biggl(\frac{2\omega_D}{\Delta}\biggr), \quad {\rm for
\quad\bar s <\Delta}\\
&=& \ln\biggl(\frac{2\omega_D}{\Delta}\biggr) - {\rm cosh}^{-1}\bigg(\frac{\bar s}{\Delta}\biggr) + \sqrt{1-\biggl(\frac{\Delta}{\bar s}\biggr)^2} ,\nonumber\\
&& \qquad\qquad\qquad {\rm for\quad  \bar s >\Delta}
\end{eqnarray}
\begin{figure}[ht]
\begin{picture}(250,200)
\leavevmode\centering\includegraphics{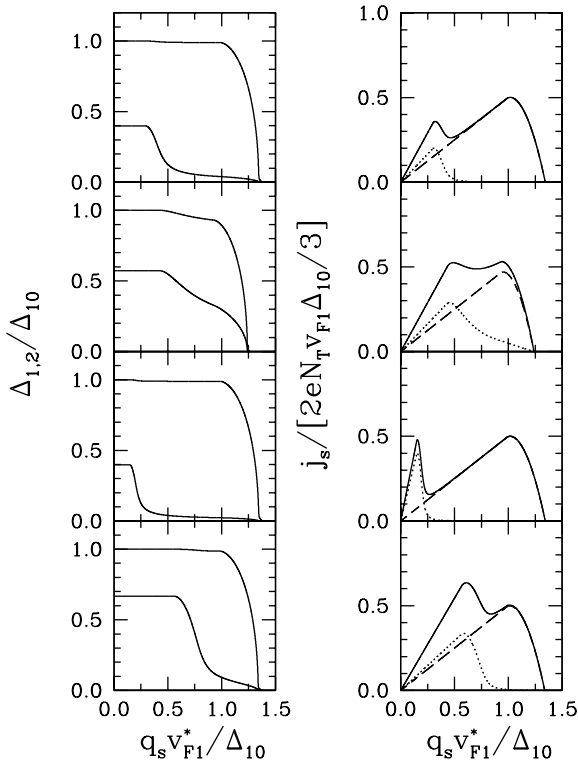}
\end{picture}
\vskip 70pt
\caption{Renormalized BCS calculations for T=0 showing, on the
left, the upper and lower gaps as a function of the superfluid
momentum $q_s$.
The top frame is for $\lambda_{11}=1.0$,
$\lambda_{22}=0.5$ and $\lambda_{12}=\lambda_{21}=0.01$, with
$v_{F2}/v_{F1}=1.0$. In the second,
 the interband coupling $\lambda_{12}=\lambda_{21}$
has been changed to 0.1. In the third frame, the parameters
are the same as for the first frame but now $v_{F2}/v_{F1}=2.0$. Finally,
for the bottom frame, the parameters are the same as for the top
frame but now $\lambda_{22}=0.7$. On the right is shown the total
superfluid
current density $j_s$ for each case (solid line) and those components for the
first band $j_{s1}$ (dashed line) and and the second band $j_{s2}$ (dotted line).
}
\label{fig1}
\end{figure}
These equations need to be solved numerically.
Results are shown in the four lefthand panels of Fig.~\ref{fig1}. The
top frame is for 
 $\lambda_{11}=1$, $\lambda_{22}=0.5$,
$\lambda_{21}=\lambda_{12}=0.01$ (nearly decoupled bands)
and $v_{F1}=v_{F2}$. Because $\lambda_{12}$ is so small, the larger
gap $\Delta_1$ exhibits the known behavior as a function of 
$q_sv_{F1}^*/\Delta_{10}$ of a one-band s-wave
superconductor\cite{maki} with a small modification.
In one-band, 
the gap
would be unchanged until $q_sv_{F1}^*/\Delta_{10}\simeq 1.0$  
at which point it would begin a sharp drop toward zero which is
reached at $q_sv_{F1}^*/\Delta_{10}\simeq 1.35$. Here, the same
behavior
occurs but there is a very slight drop from the $q_s=0$ value of
$\Delta$ at $q_sv_{F2}^*/\Delta_{20}\simeq 1.0$ due to the coupling
to the second band.
The lower gap behaves differently. It retains its $q_s=0$
value until $q_sv_{F2}^*/\Delta_{20}\simeq 1.0$ at which point it also
begins a rapid drop but retains a small finite value until the point when 
$\Delta_1=0$. This small tail is due to the coupling 
 $\lambda_{12}=\lambda_{21}$ which guarantees that the second gap
is nonzero, although it can be very small, as long as the first gap is finite.
In the second frame  $\lambda_{12}=\lambda_{21}$ has been increased to
0.1. This results in a considerable integration of the two gaps.
At $q_s=0$ the lower gap ratio $\Delta_2/\Delta_{10}$ has increased
from the 0.4 value of the top frame to 0.57. More significantly for the present
paper, as $q_s$ increases both gaps retain their $q_s=0$ value until
$q_sv_{F2}^*/\Delta_{2}\simeq 1.0$, which is the point at which the 
second gap starts to decrease, but this drop is not as fast as in the
top
frame because of the large interband coupling. At the same point
as the second gap starts to decrease, the first is also reduced but by 
much less. Beyond $q_sv_{F1}^*/\Delta_{10}\simeq 1.0$ both
gaps drop sharply as expected and reach zero together at 
$q_sv_{F1}^*/\Delta_{10}\simeq 1.27$, a value which is 
smaller than the one band value of 1.35. 
This shift in zero along the horizontal
axis is also accompanied by a shift to the left,
relative to the decoupled case, of the inflection points just
described, but these shifts are small.
These are signatures of the
two-band
nature of our system. In the third frame of Fig.~\ref{fig1} (lefthand
side), we have retained  $\lambda_{12}=\lambda_{21}=0.01$ but now have
increased the ratio $v_{F2}/v_{F1}$ to 2.0  instead of 1.0. 
This has the effect of reducing the region of $q_s$ in which the 
second gap remains constant at its $q_s=0$ value and is due
to the increase in $v_{F2}^*$. For the last frame, we have kept 
the parameters as in the first frame but have increased $\lambda_{22}$
to a 
value of 0.7. This has the effect of increasing the $q_s=0$ value of
the
second gap and also increasing the range over which it
stays
constant. Additionally, the tail beyond
$q_sv_{F2}^*/\Delta_{20}\simeq 1.0$
is not as small as it is in the top frame due to less gap anisotropy.

\begin{figure}[ht]
\begin{picture}(250,200)
\leavevmode\centering\includegraphics{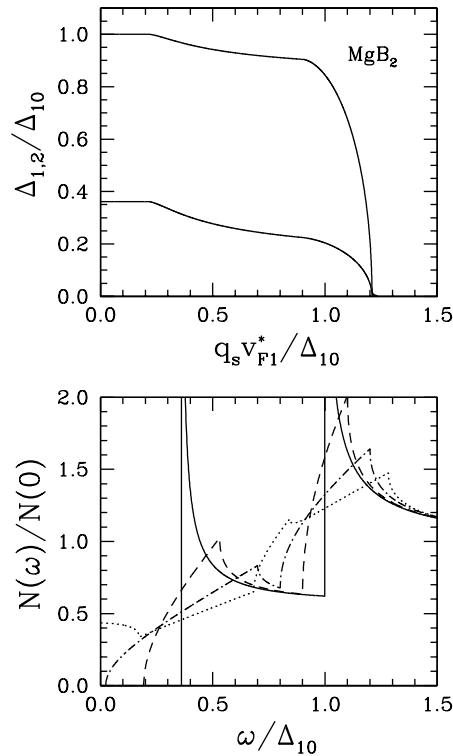}
\end{picture}
\vskip 70pt
\caption{Upper frame: The two gaps for MgB$_2$ parameters,
calculated at T=0 for in the renormalized BCS theory, shown as
a function of the superfluid momentum $q_s$. Lower frame:
The density of states $N(\omega)/N(0)$ as a function of $\omega$
is shown for several values of $q_sv^*_{F1}/\Delta_{10} = 0.0001$ (solid),
0.1 (dashed), 0.2 (dot-dashed) and 0.3 (dotted).
}
\label{fig2}
\end{figure}

In renormalized BCS, Eq.~(\ref{eq:rbcsjs}) for the partial current
can be evaluated analytically at zero temperature to obtain
\begin{eqnarray}
j_{si}(0)&=& \frac{en_i}{1+\lambda_{ii}+\lambda_{ij}}q_s, \qquad {\rm for}\quad 
\frac{\bar s_i}{\Delta_{i0}}<1\\
&=& \frac{en_i}{1+\lambda_{ii}+\lambda_{ij}}q_s\biggl[1-\biggl[1-\biggl(
\frac{\Delta_{i0}}{\bar s_i}\biggr)^2\biggr]^{3/2}\biggr],\nonumber\\
&& \qquad\qquad\qquad \qquad {\rm for} 
\frac{\bar s_i}{\Delta_{i0}}>1
\end{eqnarray}
Results for the current are presented in the righthand side frames of
Fig.~\ref{fig1}
for the four sets of parameters considered 
on the lefthand side. In all cases, the dotted lines give the results
for the partial contribution to the current coming from the second
band and the dashed for those from the first. The sum of the two partial
contributions  is the solid curve. A first feature to note is that
for the three frames for which 
$\lambda_{12}=\lambda_{21}=0.01$ (nearly decoupled),
the dashed curves are essentially identical. The straight line
segment has slope 0.5 in our units and the zero is very nearly
at $q_sv_{F1}^*/\Delta_{10}= 1.35$ as in the one-band case.
Note that on the vertical axis, we have used the total  electronic
density of states $N_T=N_1(0)+N_2(0)$ with $N_1(0)=N_2(0)$.
If we had used only $N_1(0)$ then the slope of the straight line
segment would have been 1 which corresponds to what is expected in the
one-band case. A second point to note is that increasing 
$\lambda_{12}=\lambda_{21}$ to 0.1 as in the second frame from the top 
has not had a large effect on the partial current from the first band
except to reduce somewhat the value of $q_s$ at which the current 
goes to zero. This is accompanied by a small shift to the left in
the position of the peak and a small reduction in its height.
By contrast, the effect of the parameter changes are
much more significant on the dotted curve (the current in the
second band). In the top frame, the slope of the straight line
portion of the dotted curve is a little larger than 0.5
because of the renormalization factors on $v_{F2}^*$
relative to $v_{F1}^*$. More
significantly the partial current starts to deviate from linearity
at $q_sv_{F2}^*/\Delta_{20}\simeq 1$ at which point it drops
rapidly towards zero but remains finite (although small) until the
current in the first band has reached zero. This also applies to the
third frame although  in that case the current in the second band has
been  increased  in value because of the increase in $v_{F2}$
by a factor of two. In all cases the total current (solid curve)
exhibits two peaks. The peak arising at lower values of $q_s$ has 
a significant contribution from both bands while the other
peak at higher values of $q_s$ is due almost
entirely to the partial current  of the first band when
$\lambda_{12}=\lambda_{21}$ is very small. On the other hand, when there is
more
integration of the two systems (second frame from top) both
bands can make a significant contribution. In the top three frames,
the first peak is lower than the second while in the fourth frame, 
the opposite holds. The height of the first peak can be made
higher
by increasing the value of $v_{F2}$ as in the third frame compared
with the first (or equivalently, increasing $N_2/N_1$
as the density of states enters in combination with $v_F$ and so
will have a similar effect, which is redundant to discuss here)
or by increasing $\lambda_{22}$ as in the fourth,
again compared with the first. It can also be increased by
increasing the offdiagonal coupling as in the second frame. When this
is done, however, the contribution of the second band to the total
current  in the second peak is also increased.
Finally we note that the peak heights in the two middle frames
are only very slightly different and the lower $q_s$ peak could
be made to be the highest through a small tuning of the parameters.
As the highest peak determines the value of $j_c$, this transfer from
one peak to the other by varying parameters is very interesting
and will produce nonstandard effects at finite temperature. 

\subsection{RBCS at T=0 for MgB$_2$}
 
Before leaving our discussion of RBCS at $T=0$,
it is of interest to consider the specific case of MgB$_2$
now well-established to be a two-band electron-phonon
superconductor. Microscopic parameters have been calculated from
extensions of bandstructure calculations to   obtain the
electron-phonon
spectral densities. The Coulomb repulsion parameters that enter
the gap equations (\ref{eq:Del}) and (\ref{eq:Z})
are also known. The parameters are
 $\lambda_{\sigma\sigma}=1.017$, $\lambda_{\pi\pi}=
0.448$, $\lambda_{\sigma\pi}=0.213$, $\lambda_{\pi\sigma}=0.155$,
$\mu^*_{\sigma\sigma}=0.210$, $\mu^*_{\pi\pi}=0.172$, $\mu^*_{\sigma\pi}
=0.095$, $\mu^*_{\pi\sigma}=0.069$, with $\omega_c=750$ meV. The ratio
of the two density of states $N_\pi(0)/N_\sigma(0)= 1.37$ and of the Fermi
velocities $v_{F\pi}/v_{F\sigma}=1.2$. To switch to our notation,
$(\sigma,\pi)=(1,2)$.
We start with RBCS results. In the top frame of Fig.~\ref{fig2},
we show the dependence of $\Delta_1$ and $\Delta_2$ normalized to
$\Delta_{10}$ as a function of $q_sv_{F1}^*/\Delta_{10}$. On
comparison
with some of the results of Fig.~\ref{fig1}, we note that
MgB$_2$ shows considerable integration of the two bands and also 
both gaps become zero at  $q_sv_{F1}^*/\Delta_{10}\simeq 1.2$
which is considerably reduced from the one-band BCS value. 

\begin{figure}[ht]
\begin{picture}(250,200)
\leavevmode\centering\includegraphics{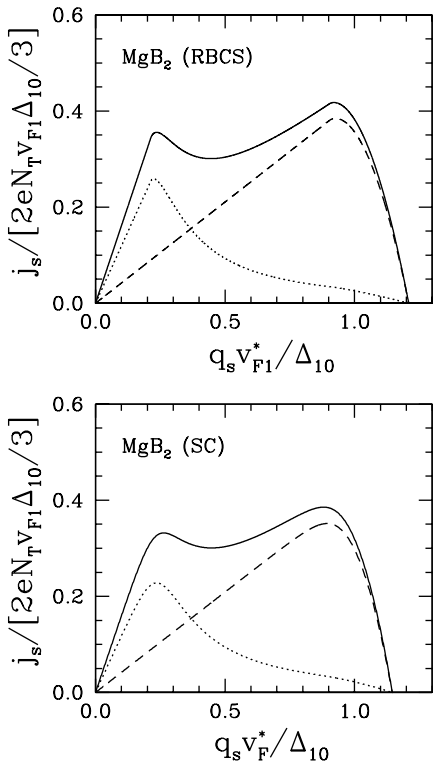}
\end{picture}
\vskip 70pt
\caption{The current $j_s$ as a function of $q_s$ calculated for
the MgB$_2$ parameters. Shown are the component contributions
of the currents for the first and
second bands (dashed and dotted curves, respectively) and the total
current (solid  curve). The upper frame shows the calculation for
$T=0$ RBCS and the bottom frame shows the curves calculated in
Eliashberg theory for $T/T_c=0.1$.
}
\label{fig3}
\end{figure}

An
interesting
quantity that could be measured is the quasiparticle density of 
states $N(\omega)/N(0)$ in the presence
of a current, which was first examined for
one-band s-wave superconductors by Fulde\cite{fulde}. 
Such measurements have recently been
reported for Pb in Ref.~\cite{pbref}.
For each band separately
\begin{eqnarray}
N_i(\omega)&=&\frac{N_i(0)}{2\bar s_i}\Re e\biggl[\frac{\omega+\bar
    s_i}{|\omega+\bar s_i|}\sqrt{(\omega+\bar s_i)^2-\Delta_i^2}\nonumber\\
&&-
\frac{\omega-\bar
    s_i}{|\omega-\bar s_i|}\sqrt{(\omega-\bar
    s_i)^2-\Delta_i^2}\biggr]
\end{eqnarray}
with $N(\omega)=N_1(\omega)+N_2(\omega)$. In the bottom
frame of Fig.~\ref{fig2}, we give results for $N(\omega)/N_T(0)$
as a function of $\omega/\Delta_{10}$ for various values of 
$q_s$ as indicated in the figure. The solid curve is essentially
the zero current limit and shows the expected superposition of 
two BCS quasiparticle density of states with square root
singularities at the two gaps. As the current is increased in
the linear region only (cross-referencing with the upper
frame of Fig.~\ref{fig3} which shows $j_s$
for MgB$_2$), the singular structure is smeared 
and the gap region fills as does the region between the two
gaps. Gapless behavior appears at a value of $q_s$
near the first peak in the total current 
due to the density of states in
the small band going gapless. This is in contrast to what is found
in a one-band s-wave superconductor\cite{fulde} where the 
gapless behavior occurs at the critical current.

The superfluid current $j_s$ versus $q_s$ for the RBCS calculation
with MgB$_2$ parameters is plotted in the upper frame of
Fig.~\ref{fig3}.
The line types are defined in the same way as for the $j_s$ curves
in Fig.~\ref{fig1}. We note that this graph is very similar to that
shown in the second frame (righthand side) of Fig.~\ref{fig1}, where
it has
been discussed in relation to other cases. As   in that graph, the
peak in the total current seen at low $q_s$ is lower than the one
at higher $q_s$, so that
this second peak determines the critical current.
 It is of considerable interest to compare these 
renormalized BCS results with those from full Eliashberg calculations
based on the $\alpha^2_{ij}F(\omega)$ spectra given in
Ref.~\cite{goliop}. Very similar values of the parameters can
also be found in Ref.~\cite{cohen}.
Such results are presented  in the lower frame of Fig.~\ref{fig3}.
There are several differences that are worthy of note. The full
Eliashberg results are at $T/T_c=0.1$ and the RBCS are for 
$T=0$. This difference in temperature is small and is
not expected to lead to any  significant differences.
Examination of the partial currents shows that 
 the
dashed curves are almost the same in both plots except that
strong coupling
effects have shifted the point of zero current to a slightly smaller
value of $q_s$. 
This is accompanied by a shift to the left of the
position of the peak,  a reduction in its height and an 
increase in the rounding, all small effects. 
These differences are seen to be
slightly more prominent  in the dotted curve. This
shows that RBCS can be used with confidence as a first
approximation.
However,
from this point on, we will consider only Eliashberg results
as we have available to us the electron-phonon spectral functions
and we wish to consider the full effects of temperature and impurities.

\begin{figure}[ht]
\begin{picture}(250,200)
\leavevmode\centering\includegraphics{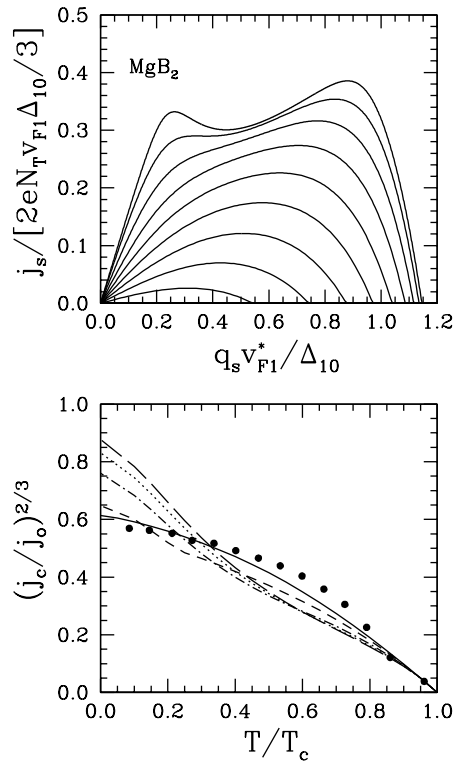}
\end{picture}
\vskip 70pt
\caption{Upper frame: The temperature evolution of the $j_s$
versus $q_s$ for MgB$_2$ calculated in Eliashberg theory.
From top to bottom, the temperatures range from $T/T_c=0.1$ to 0.9  in
steps of 0.1.
Lower frame:
The critical current plotted as $j_c^{2/3}$ versus $T/T_c$ for MgB$_2$ with
varying $v_{F2}/v_{F1}$ ratio equal to 1.2 (solid curve -
corresponding
to the set of curves shown in the upper frame), 2 (short dashed), 3
(dot-dashed), 4 (dotted), and 5 (long-dashed). The solid dots are the data
for MgB$_2$ 
taken from Kunchur\cite{kunchur}. The data appears to be in good agreement
with the
solid curve which corresponds to the standard set of parameters given
for MgB$_2$ in the literature.
}
\label{fig4}
\end{figure}

\section{Eliashberg calculations for finite $T$ and impurities}

\begin{figure}[ht]
\begin{picture}(250,200)
\leavevmode\centering\includegraphics{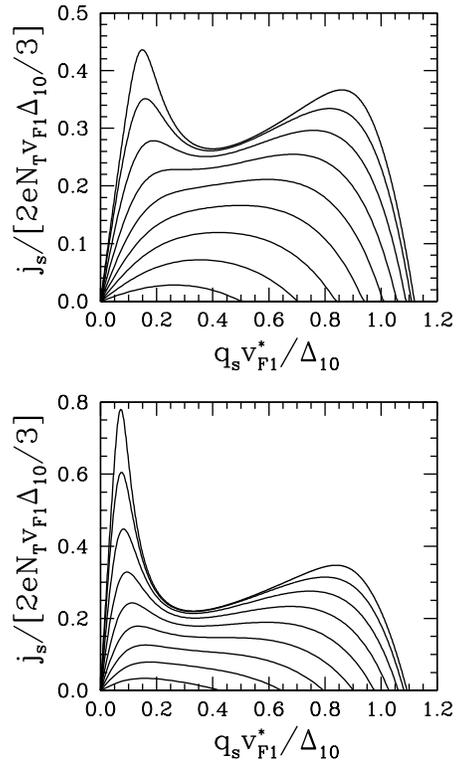}
\end{picture}
\vskip 70pt
\caption{The temperature evolution of the $j_s$
versus $q_s$ curves for MgB$_2$ calculated in Eliashberg theory.
From top to bottom, the temperatures range from $T/T_c=0.1$ to 0.9  in
steps of 0.1. The upper frame is for a 
$v_{F2}/v_{F1}$ ratio equal to 2
and the lower frame is for a ratio  of 4.
}
\label{fig5}
\end{figure}

At the end of the last section, we showed a comparison between
RBCS and Eliashberg theory (Fig.~\ref{fig3}). Given that the
electron-phonon spectral functions for MgB$_2$ are available,
it is appropriate to work with the full theory 
 for $j_c$ as a function of temperature and
impurity scattering.
We begin with a consideration of temperature effects. In the upper
frame of Fig.~\ref{fig4}, we show the current $j_s(T)$ normalized by
$2eN_Tv_{F1}\Delta_{10}/3$ as a function of $q_sv_{F1}^*/\Delta_{10}$
for nine values of temperature from $T/T_c=0.1$ to  0.9 in steps of
0.1.
The parameters used are those appropriate to MgB$_2$ with the upper
curve reproduced from the bottom frame of Fig.~\ref{fig3}.
As the temperature is increased, the peak with larger value of $q_s$
remains the highest and therefore 
determines the critical current given
as the solid curve in the lower frame. Here,
$j_c$ has been normalized by $j_0$ so that the slope at $t=1$
is negative one by arrangement for the typical critical current
plot of $j_c^{2/3}$, which is designed to bring out the
Ginzburg-Landau
behavior near $T_c$.
We see a smooth (concave 
downward) increase in $j_c(t)$ as $t$ is reduced. The value
of the critical current at $t=0$ is $\sim 0.61$. This is 
lower than the one-band BCS value of 0.72. This reduction could reflect
the two-band nature of MgB$_2$ as well as strong coupling
corrections\cite{nicolmg,nicoljc}
due to our use of the full Eliashberg equations
(\ref{eq:Del}) and (\ref{eq:Z})
to calculate the current
given by Eq.~(\ref{eq:js2}). However,
reference to the lower frame of Fig.~\ref{fig3} shows
that the second band contributes only about 10\% to the total critical
current at $T=0$ and this remains true at $T_c$. Consequently, this
curve is very nearly the single band result and deviates from the
classical BCS curve almost entirely because of retardation effects.
Also shown is the data of Kunchur\cite{kunchur}, plotted as solid dots.
There is good agreement between the data and the solid curve which
is quite remarkable as the calculation has no free parameters, in principle,
as they are all given in the literature.
Including impurity scattering has an effect on the temperature
dependence of $j_c(t)$, 
as we will discuss later in relation to Fig.~\ref{fig7}.
Unfortunately, information 
on the individual intraband scattering rates in these films for
$\pi$ and $\sigma$ bands, separately, is not available to us. Including
some impurity scattering in the $\pi$ band would  futher improve the
agreement with experiment.
The other critical current curves in Fig.~\ref{fig4} are for different
ratios of $v_{F2}/v_{F1}$ as indicated in the figure caption.
In addition to allowing for some uncertainty in the reported $v_F$
values in the literature, this also mimics the possibility of currents
associated with the c-axis where the $\sigma$-band is known to have a 
much smaller $v_F$ than the $\pi$-band. Likewise, we have not included
the
details of the full Fermi surface average in the expression for the
current, however, it is known that the $\sigma$-band is
quasi-two-dimensional
as opposed to the more three-dimensional $\pi$-band. Dahm and
Scalapino\cite{dahm}
find significant geometrical corrections associated with such
averaging
which could also accentuate the difference between the two bands.
This may also be approximately captured by our use of different
$v_F$ ratios.
All the curves are normalized  to have a slope of negative one at
$t=1$. As $v_{F2}$ is increased, the normalized critical current at zero
temperature
increases and can become considerably  larger than its one-band
BCS value. There is also a change in the temperature variation
of the curves, the dashed curve shows a clear kink around $t\simeq
0.25$ while the long-dashed curve shows concave upward behavior at
intermediate temperatures. This can be understood better when the
pattern of behavior shown in Fig.~\ref{fig5} is considered. 
The top
frame gives the temperature evolution for the case $v_{F2}/v_{F1}=2$
and the bottom is for a ratio of 4. The notation is as for the
top frame of Fig.~\ref{fig4} and for the same nine reduced
temperatures.
 In the top frame, it is the peak at lower $q_s$
which determines the critical current at $t=0$ while  at
higher $t$, it is the peak with higher $q_s$. This crossover 
occurs between $t=0.2$  and $t=0.3$ and leads 
to the kink seen in   the dashed curve of Fig.~\ref{fig4} (lower
frame). If we look at the individual contributions from each
band (not shown here) to the total current, we find that at $T=0$
it is the second band which makes the dominant contribution
because of its large value of $v_F$ but near $T_c$ this contribution
has decayed sufficiently that it only makes a 30\% contribution to the
total.

In the lower frame of Fig.~\ref{fig5}, 
the peak at smaller value of $q_s$ has greatly increased in magnitude
over its value in the top frame. It has also moved to
a lower value of $q_s$ as expected from Fig.~1. However, now
there is no
crossover
from smaller to larger $q_s$ peak and hence no kink. 
Nevertheless, the dotted curve
is very different from a one-band BCS variation.
We can understand better how this can arise by  once more
looking at the partial contribution to the critical current
from each band.
At $T=0$ it is the second band which
dominates while at $T_c$ it accounts only for 2/3 of the total with
the remaining 1/3 contribution from band one. This difference in
admixture of the two bands with temperature  is
seen to be enough to cause an upward curvature in the dotted
curve at intermediate temperatures as well as to increase
the value of the normalized critical current at $T=0$ to a value
considerably larger than the BCS value of 0.72. Increasing
$v_{F2}$ further as in the long-dashed curve of the lower frame
of Fig.~\ref{fig4} increases the $T=0$ value even further as the peak
at
lower $q_s$ value becomes even more prominent.
Details of the temperature variation of the critical current
clearly depend on details of the microscopic parameters
of the two bands involved, such as values of $\lambda_{ij}, N_i,
v_{Fi}$. For example, it is possible to shift the kink in $j_c(t)$
to higher values of reduced temperature and to make it sharper by
considering 
other parameter sets.
\begin{figure}[ht]
\begin{picture}(250,200)
\leavevmode\centering\includegraphics{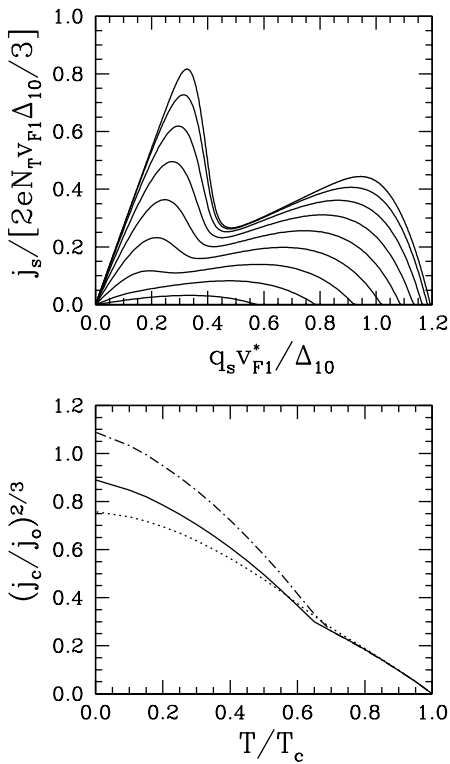}
\end{picture}
\vskip 70pt
\caption{Upper frame: The temperature evolution of curves for 
$j_s$ versus $q_s$ for
a Lorentzian spectrum with
$v_{F2}/v_{F1}= 2$, $\lambda_{11}=1$, $\lambda_{22}=0.8$,
and $\lambda_{12}=\lambda_{21}=0.01$. 
From top to bottom, the temperatures range from $T/T_c=0.1$ to 0.9  in
steps of 0.1.
Lower frame: The critical current as a function of temperature 
corresponding to the same curves in the upper frame (solid curve) and
also for the cases of 
$v_{F2}/v_{F1}= 1$ (dotted) and 3 (dot-dashed).
}
\label{fig6}
\end{figure}
To illustrate this,
in Fig.~\ref{fig6} we show results for a Lorentzian $\alpha_{ij}^2F
(\omega)$ spectrum defined  in detail in our previous
paper\cite{nicolmg}.
In  this model, $\lambda_{11}=1$ and we take $\lambda_{22}=0.8$ to
decrease
the gap anisotropy as compared with MgB$_2$.
We have also taken $\lambda_{12}=\lambda_{21}=0.01$ (nearly decoupled
bands). The top frame gives the current $j_s(t)$ for the case
of $v_{F2}/v_{F1}=2$,
for the nine 
values of reduced temperature previously chosen. With this $v_F$
ratio, 
the peak with lower
value of $q_s$ remains highest at small $t$ and the crossover of the 
maximum to the other peak occurs at a
reduced temperature between 0.6 and 0.7.
This manifests itself as a sharp kink in the solid curve
for $j_c$ (bottom frame) at this same
temperature. Also shown in this same frame are results for the
critical current when $v_{F2}/v_{F1}=1$ (dotted) and $v_{F2}/v_{F1}=3$
(dot-dashed). 
For the larger value of $v_{F2}$, the normalized current
at $T=0$ is increased over the $v_{F2}/v_{F1}=2$ case as expected
and for $v_{F2}=v_{F1}$ it is reduced but remains above BCS. In this
latter case, the two peaks in the current versus $q_s$ are closer together
than they are in the top frame of Fig.~\ref{fig6} and seem to 
almost merge before the crossover from lower to higher $q_s$ peak
occurs, which means that the kink in the dotted curve can hardly be
seen.
It is clear that a
 rich pattern of behavior exists
for the current versus $q_s$ at different
reduced $t$ in a two-band model. This translates into quite
distinct temperature dependence for  $j_c(t)$ as compared with the 
one-band BCS canonical behavior.

Some addition insight into this complicated behavior can be
obtained from a consideration of simplified but analytic
BCS results in the
limiting case of two decoupled, well-separated bands, i.e.,
for small values of the gap anisotropy $u=\Delta_{20}/\Delta_{10}$.
Under this assumption, $1/\chi_1=1$, $1/\chi_2=0$,
$1/\chi_1'=v_{F1}^{*2}$, and
$1/\chi_2'=0$ so that the critical current near $T_c$ given by
formula (\ref{eq:bcs14}) reduces to   
\begin{equation}
j_c(t)=\frac{16}{9}\frac{e\pi T_c}{\sqrt{7\zeta(3)}}
N_1v_{F1}(1-t)^{3/2},
%\label{eq:bcsa}
\end{equation}
and is determined entirely by band one, as previously commented upon.
As the temperature is lowered, however, two possible circumstances
can arise. The critical current can remain determined by the same
band (number one)  in which case we can write approximately 
\begin{equation}
j_c(0)\simeq \frac{n_1e}{m_1v_{F1}}\Delta_{10}.
\end{equation}
A second possibility is that a crossover to the peak at lower $q_s$
 value 
occurs and we have approximately
\begin{equation}
j_c(0)\simeq \biggl(\frac{n_1}{m_1}+\frac{n_2}{m_2}\biggr)
\frac{e\Delta_{20}}{v_{F2}}.
\end{equation}
For the first case, 
\begin{equation}
\biggl(\frac{j_c(0)}{j_{0}}\biggr)^{2/3}=0.72
\label{eq:bcsd}
\end{equation}
which is the one-band BCS result where we have normalized the
critical
current in such a way that $j_c^{2/3}$ has  a slope of negative one
at $t=1$. This conforms with what we have done in  the lower
frames of Figs.~\ref{fig4} and \ref{fig6}. For the second
case, by assumption, there is a crossover and 
\begin{equation}
\biggl(\frac{j_c(0)}{j_{0}}\biggr)^{2/3}=0.72\biggl[\frac{\Delta_{10}}{1.76T_c}
u\biggl\{\frac{N_2v_{F2}}{N_1v_{F1}}+\frac{v_{F1}}{v_{F2}}\biggr\}\biggr]^{2/3}.
\label{eq:bcse}
\end{equation}
This simplified approximate formula shows clearly that the anisotropy
$u$ on its own reduces the value of $(j_c(0)/j_{0})^{2/3}$ below  its
universal BCS value of 0.72 while a large value of $N_2$ relative to $N_1$
or $v_{F2}$ relative to $v_{F1}$ increases it and that this reduced
dimensionless quantity is no longer universal. It can depend on the
band
structure parameters as well as on the gap anisotropy. 

Next we consider the effect of impurities on the current. It is
instructive
to begin by expanding Eq.~(\ref{eq:js2})  for $j_s(t)$  to lowest
order
in $q_s$. Doing so leads to the well-known and physically expected
result  that for small superfluid velocity
\begin{equation}
j_s(t)=\sum_{i=1}^2j_{si}=\sum_{i=1}^2en_{si}v_s
\end{equation}
where $n_{si}$ is the superfluid density of the $i$'th band given for
a strong coupling superconductor by
\begin{equation}
n_{si}=n_i\biggl[2\pi T\sum_n\frac{\tilde\Delta_i^2(n)}{[\tilde\omega_i^2(n)
+\tilde\Delta_i^2(n)]^{3/2}}\biggr]
\label{eq:bcsg}
\end{equation}
In BCS at $T=0$, this becomes
$n_{1}/(1+\lambda_{11}+\lambda_{12})$
and 
$n_{2}/(1+\lambda_{22}+\lambda_{21})$
for band 1 and band 2, respectively. The slope of $j_s(t)/e$
as a function of $v_s$ gives $n_{s1}+n_{s2}$ with Eq.~(\ref{eq:bcsg})
valid for any temperature and any impurity content. The 
impurities enter only
through
the Eliashberg gap equations (\ref{eq:Del}) and (\ref{eq:Z}). As the
temperature is increased, the superfluid density is reduced and so the
slope of $j_s(t)$ seen in Figs.~\ref{fig4} to \ref{fig6} is
correspondingly
reduced. Impurities also reduce the superfluid density. In
BCS, the known result for each of the two bands
separately, to
lowest order, is 
\begin{equation}
n_{si}=\frac{n_i}{1+\lambda_{ii}+\lambda_{ij}}\frac{1}{1+\frac{\pi^2\xi_{oi}}
{8l_i}}
\end{equation}
with $\xi_{0i}=v_{Fi}/(\pi\Delta_{i0})$ and $l_i=v_{Fi}\tau_{ii}$,
the coherence length and impurity mean free path in the $i$'th
band, respectively.
This applies to intraband scattering which reduces the initial slope
of the partial current coming from each band.
\begin{figure}[ht]
\begin{picture}(250,200)
\leavevmode\centering\includegraphics{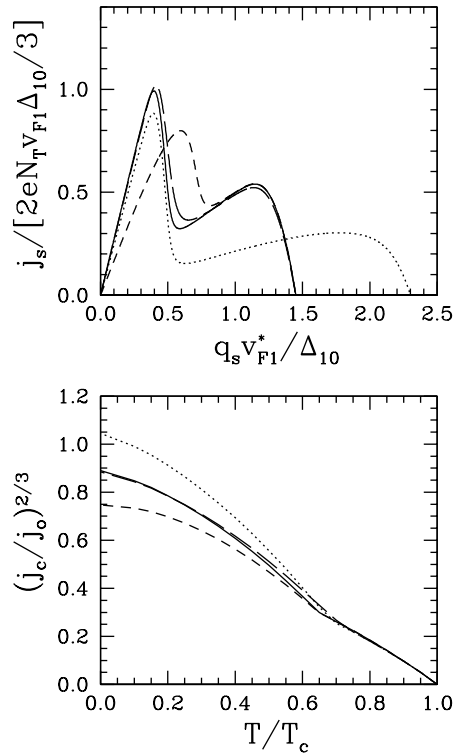}
\end{picture}
\vskip 70pt
\caption{Upper frame: The superfluid current $j_s$ versus
$q_s$ for a Lorentzian spectrum as used in 
Fig.~\ref{fig6}, where the solid
curve is repeated from Fig.~\ref{fig6} for $T/T_c=0.1$
and has a ratio of the
two $v_F$'s equal to 2. The dotted curve is where a scattering
rate of $t^+_{11}=2T_{c0}$ is now introduced, the short dashed is
for $t^+_{22}=T_{c0}$, and the long-dashed is for
$t^+_{12}=t^+_{21}=0.02T_{c0}$. Lower frame:
The critical current as a function of temperature corresponding to the
parameters for the curves shown in the upper frame.
}
\label{fig7}
\end{figure}
Results are presented in Fig.~\ref{fig7}, where we also consider
the interband case $t^+_{ij}\ne 0$. In the top
frame, we show $j_s$  
versus $q_s$ for the reduced temperature $t=0.1$
in the Lorentzian model used in  Fig.~\ref{fig6}. The solid curve applies to
the
pure case and is for comparison. We show only three cases,
one where $t^+_{11}=2T_{c0}$ (dotted curve), $t^+_{22}=T_{c0}$ (dashed
curve), 
and $t^+_{12}=t^+_{21}=0.02T_{c0}$, where
$T_{c0}$ is the $T_c$ without impurities (in the case
of $t^+_{11}$ and $t^+_{22}$, $T_c$ will be unaffected, but
not so for interband impurities). Intraband impurities
reduce only the partial contribution to the current 
coming from that band. It reduces its value on the vertical
axis and extends the x-axis to higher values of $q_s$
leaving the second band largely unaffected. Of course, since both
bands contribute significantly to the total current in the region
of the peak corresponding to the smaller value of $q_s$, its
magnitude is reduced even in the dotted curve. On the other hand,
for the dashed curve, there is little change in the region
of the second peak which is dominated by the first band.
By contrast interband impurities modify both bands in
a similar fashion. The corresponding critical current  as a function
of $T/T_c$ is shown in the lower frame. The solid line is the pure
case
of Fig.~\ref{fig6}. As expected, a finite value of $t^+_{11}$
has reduced the critical current below its value in the pure case
at all temperatures, however,
when normalized  to a slope of magnitude one near $T_c$, the effect
pushes the curve up at low $T$. A kink remains and the crossover temperature
is only slightly shifted toward higher values. For the dashed curve,
the significant modifications arise only below the kink
temperature. By contrast the main effect of the interband
impurities is to wash out the kink structure leaving the rest of the
curve largely unaffected (of course the $T_c$ would show a small
shift downwards relative to $T_{c0}$, not shown here).

\section{Conclusions}

In conventional one-band s-wave superconductors, the energy gap
$\Delta$ remains unaffected by the flow of a supercurrent up to a
critical value of the superfluid velocity $v_s=q_s/m$
(ie., $v_s^c=\Delta/mv_F$) 
at which point
pairbreaking becomes possible. In this region, the current $j_s$
is linear in $v_s$ with proportionality constant equal to 
the superfluid density. Beyond this critical value of $v_s$,
the rapid reduction in the gap in addition to a quasiparticle
backflow current leads to $j_s=0$ at $\Delta=0$. For 
two nearly decoupled bands, we would expect that
the total current to
 exhibit a two-peak structure as a function
of $v_s$. A first one near the  critical value of the superfluid
velocity for band two and the other near the larger critical
value of $v_s$ corresponding to the first band alone. The first peak
has in general a significant contribution from both bands
(unless $v_{F2}\gg v_{F1}$)
while the second peak could be due mainly to the first band. The
relative admixture depends on microscopic parameters such as
density of states $N_i(0)$ and Fermi velocities $v_{Fi}$,
as well as gap anisotropy. 
When $v_{F2}$ is made large as compared to $v_{F1}$,
the first peak moves to lower values of $v_s$ and can also become larger
than the second at low temperature. As the temperature is increased
the superfluid density in each band is reduced but that in the
second band is more strongly affected because of its smaller gap value.
Consequently, since the critical
current is determined by the height of the highest peak, one can have
a crossover from the first to second peak as $T$ is raised from
$0$ to $T_c$ and this will reflect itself as a kink in the temperature
dependence of the critical current. 
Such a feature is also found in the dirty limit calculations presented in
Ref.~\cite{koshelevjc} for the critical current along the c-axis.
When there is a significant
interband
coupling or the two gaps are close in value (isotropic case) the
two peaks in the total current cease to be independent and a rich
pattern of behavior emerges for the temperature and impurity
dependence of the two peaks. In particular, as the temperature is
raised, the first peak can remain the
determining peak for $j_c$ up to $T_c$ or it can become
the smaller of the two at a crossover temperature which can be 
increased or decreased through judicious  choices of parameters.
Another possibility is that the two peaks merge into one or that
the second peak is the dominant  one at all temperatures.
Application to the specific case MgB$_2$, which involves no adjustable
parameters, gives good agreement with the in-plane data of 
Kunchur\cite{kunchur}.

If the resulting critical current $j_c^{2/3}$ is normalized to
have slope of negative 
one as a function of $(1-t)$ near $t=1$, the corresponding
value of the critical current at $T=0$ will differ from its classic
one-band BCS value of 0.72. Strong coupling effects resulting from our
use
of the full Eliashberg equations, with  appropriate electron-phonon
spectral densities describing the electron-phonon interaction,
are known\cite{nicoljc} to always
reduce the normalized value of $j_c(0)$ while two-band effects can
reduce  or increase it depending on the detailed values of the
microscopic parameters  involved.  In particular, a large value of
$v_{F2}$
relative to $v_{F1}$ can lead to values that can be even larger
than one. This is expected for the case of current orientated along
the c-axis for the specific parameters representative of 
MgB$_2$. In this case, because the $\sigma$ band (band one in 
our notation) is nearly two-dimensional, the Fermi velocity of the
$\pi$ band in the c-direction will be much greater than that for  the
$\sigma$ band.

The superfluid density in each  subband, which determines the initial
slope of the linear in $v_s$ regime for the supercurrent, is reduced
with the increase in intraband impurity scattering. This fact can be
used to manipulate the position along the $v_s$ axis and size of the
two peaks in the total current and consequently the temperature 
dependence of the resulting critical current. Introducing impurities
in the first band will reduce  the critical current at all
temperatures since this band always makes a  contribution
to the total current. On the other hand, if they are introduced only
in the second band, this will leave the critical current largely 
unaltered around $T=T_c$. Interband impurity scattering affects both
bands although not necessarily in exactly the same way and leads
to the smoothing out of the kink in $j_c^{2/3}(t)$ versus $t$
when it exists.

In conclusion, we predict a complex pattern of behavior for the
total current as a function of $v_s$, temperature and impurity
content which can be used to restrict further the values of the
microscopic parameters involved in two-band superconductors, such as
MgB$_2$.

\begin{acknowledgments}
EJN acknowledges funding from NSERC, the
Government of Ontario
(PREA), and the University of Guelph.
JPC acknowledges support from NSERC and the CIAR. 
This research was supported in part by the National Science
Foundation under Grant No. PHY99-07949.
\end{acknowledgments}


\begin{thebibliography}{99}

\bibitem{carbotte} J.P. Carbotte, Rev. Mod. Phys.
{\bf 62}, 1027 (1990).

\bibitem{nicolmg} E.J. Nicol and J.P. Carbotte, Phys. Rev. B {\bf 71},
054501 (2005).

\bibitem{rogers} K.T. Rogers, Ph.D. thesis, University of Illinois, 1960
(unpublished).

\bibitem{bardeen} J. Bardeen, Rev. Mod. Phys.
{\bf 34}, 667 (1962).

\bibitem{parameter} R.H. Parameter, {\it RCA Rev.} {\bf 23}, 323,
(1962);
R.H. Parameter, and L.J. Berton, {\it RCA Rev.} {\bf 25}, 596, (1964)

\bibitem{makiold} K. Maki, Progr. Theoret. Phys. (Kyoto) {\bf 29}, 10 and
333 (1963).

\bibitem{maki} K. Maki, in Superconductivity, edited by
R.D. Parks (Dekker, New York, 1969), p.1035.

\bibitem{fulde} P. Fulde, Phys. Rev. {\bf 137}, A783 (1965).

\bibitem{lukichev} M.Y. Kupriyanov and V.F. Lukichev, Sov. J. Low Temp.
Phys. {\bf 6}, 210 (1980).

\bibitem{tinkham} M. Tinkham, {\it Introduction to Superconductivity}
(McGraw-Hill, New York, 1996).

\bibitem{lemberger} T.R. Lemberger and L. Coffey, Phys. Rev. B {\bf
  38},
7058 (1988).

\bibitem{nicoljc} E.J. Nicol and J.P. Carbotte, Phys. Rev. B {\bf 43},
10210 (1991).

\bibitem{ting} D. Zhang, C.S. Ting, and C.-R. Hu, Phys. Rev. B {\bf 70},
172508 (2004).

\bibitem{hykeenodal} I. Khavkine, H.Y. Kee, and K. Maki, Phys. Rev. B
{\bf 70}, 184521 (2004).

\bibitem{hykee} H.Y. Kee, Y.B. Kim, and K. Maki, Phys. Rev. B
{\bf 70}, 052505 (2004).

\bibitem{kunchur} M.N. Kunchur, cond-mat/0409402; M.N. Kunchur, S.I. Lee,
W.N. Kang,
Phys. Rev. B {\bf 68}, 064516 (2003).

\bibitem{koshelevjc} A.E. Koshelev and A.A. Golubov, Phys. Rev. Lett.
{\bf 92}, 107008 (2004).

\bibitem{dahm} T. Dahm and D.J. Scalapino, cond-mat/0409571


\bibitem{vekhter} See for example, I. Vekhter, P.J. Hirschfeld,
and E.J. Nicol, Phys. Rev. B {\bf 64},
064513 (2001).

\bibitem{sauls} D. Xu, S.K. Yip, and J.A. Sauls, Phys. Rev. B {\bf 51},
16233 (1995).

\bibitem{goliop} A.A. Golubov, J. Kortus, 
O.V. Dolgov, 
O. Jepsen, Y. Kong, O.K. Andersen, B.J. Gibson, K. Ahn, and R.K. Kremer,
J. Phys.: Condens. Matter {\bf 14}, 1353 (2002).

\bibitem{cohen} H.J. Choi, D. Roundy, H. Sun, M.L. Cohen, and S.G. Louie,
Nature (London)
{\bf 418}, 758 (2002).

\bibitem{aniso} H.G. Zarate and J.P. Carbotte, Phys. Rev. B
{\bf 27}, 194 (1983) and {\it Anisotropy Effects in Superconductors}
edited by H. Webber (Plenum Press, New York, 1977). 

\bibitem{pbref} A. Anthore, H. Pothier, and D. Esteve, Phys. Rev. Lett.
{\bf 90}, 127001 (2003).

\end{thebibliography}
\end{document}